\documentclass[]{spie}  
\usepackage[dvips]{graphicx}

\newcommand{\dunno}{\textbf{(???)}}
\newcommand{\infocus}{InFOC$\mu$S}

\title{The \infocus\ Hard X-ray Telescope: Pixellated CZT
Detector/Shield Performance and Flight Results} 

\author{Wayne H. Baumgartner\supit{a,b}, J. Tueller\supit{a},
H. Krimm\supit{a}, S. Barthelmy\supit{a},\\ 
F. Berendse\supit{a,b}, L. Ryan\supit{a}, F. Birsa\supit{a},
T. Okajima\supit{c,a},\\
H. Kunieda\supit{d}, Y. Ogasaka\supit{c}, Y. Tawara\supit{c},
K. Tamura\supit{c}
\skiplinehalf
\supit{a}NASA Goddard Space Flight Center, Greenbelt, MD USA\\
\supit{b}University of Maryland, College Park, MD USA\\
\supit{c}Nagoya University, Nagoya, Japan\\
\supit{d}Institute of Space and Astronautical Science, Sagamihara, Japan\\
}

\authorinfo{Send correspondence to
W.~H.~Baumgartner\\
Address: NASA Goddard Space Flight Center, Code 661, Greenbelt, MD
20771, USA\\
E-mail: wayne@astro.umd.edu\\
}

\begin{document} 
\maketitle 

\begin{abstract}

The CZT detector on the \infocus\ hard X-ray telescope is a pixellated
solid-state device capable of imaging spectroscopy by measuring the
position and energy of each incoming photon.  The detector sits at the
focal point of an 8\,m focal length multilayered grazing incidence
X-ray mirror which has significant effective area between 20--40 keV.
The detector has an energy resolution of 4.0\,keV at 32\,keV, and the
\infocus\ telescope has an angular resolution of 2.2 arcminute and a
field of view of about 10 arcminutes.  \infocus\ flew on a balloon
mission in July 2001 and observed Cygnus X-1.  We present results from
laboratory testing of the detector to measure the uniformity of
response across the detector, to determine the spectral resolution,
and to perform a simple noise decomposition.  We also present a hard
X-ray spectrum and image of Cygnus X-1, and measurements of the hard
X-ray CZT background obtained with the SWIN detector on \infocus.

\end{abstract}


\section{INTRODUCTION}

The International Focusing Optics Collaboration for $\mu$-Crab
Sensitivity (\infocus) is a balloon borne hard X-ray telescope for the
hard X-ray bands 20--40 and 65--70\,keV.  \infocus\ uses grazing
incidence multilayer mirrors\cite{b,og,ok,owens} to focus astronomical
photons onto a pixellated planar CdZnTe (CZT) detector.

CZT detectors are a natural fit for high energy X-ray astronomy.  They
have better energy resolution than scintillator detectors such as CsI
or NaI, and the rather large bandgap allows us to dispense with the
cryogenic cooling necessary for Ge detectors.  The high atomic number
of the CZT constituents provides a large cross section for
photoelectric interaction with photons, allowing detectors to be built
with small thicknesses in order to reduce sensitivity to volume
dependent background components such as the particle flux in the upper
atmosphere. CZT can be manufactured with pixellated contacts that
allow fine spatial determination of the incoming photons, and when
operated at moderate temperatures (around 0$^\circ$\,C) the low
leakage current of the CZT causes minimal noise and permits a very
high sensitivity.

The complete \infocus\ design calls for four mirrors and four focal
planes.  This plan allots one mirror for the low energy band of
20--40\,keV, and three mirrors to cover the higher energy band
65--70\,keV around the 68\,keV $^{44}$Ti line from supernova remnants.
Each of the confocal 8 meter focal length mirrors will have its own
CZT detector and shield.  The design parameters for the complete
\infocus\ telescope are given in Table~\ref{parameters}.
\begin{table}[!bth]
\caption{\infocus\ parameters.  These numbers for the low energy
mirror and detector document the details of the July 2001 science flight.}
\begin{center}
\begin{tabular}{lll}
\hline \hline
Parameter & Low Energy & High Energy \\
\hline
Focal Length		&8.0\,m	& 8.0\,m\\
Bandpass		&20--40\,keV	& 65--70\,keV\\
Number of Mirrors	&1		& 3\\
Mirror Diameter		&40\,cm	& 30\,cm\\
Effective Area		&42\,cm$^2$ @ 30\,keV	& 70\,cm$^2$\\
Field of View 		&9.6\,arcmin	& 2--3\,arcmin\\
Angular Resolution 	&2.2\,arcmin FWHM	& 1\,arcmin\\
PSF 	& 4\,mm FWHM	& 2\,mm FWHM\\
\\

Detector Material	&CdZnTe (CZT)&CZT\\
Detector Size		&2.7$\times$2.7$\times$0.2\,cm\\
Pixel Array Size	&12 $\times$ 12\\	
Pixel Spacing		&2.1\,mm\\
			&54\,arcsec\\
Active Shield		&3.0\,cm CsI\\
Energy Resolution	&4.0\,keV @ 32\,keV	& 5\,keV\\
$\Delta$E/E		&8\% at 60\,keV\\
Threshold		&18\,keV\\
Shield FOV		&8.1$^{\circ}$\\
Background		&$(2.7\pm1.2)\times 10^{-4}$\\
			&cts/cm$^{2}$/sec/keV\\
\hline
\end{tabular}
\end{center}
\label{parameters}
\end{table}

\begin{figure}
\begin{center}
\resizebox{!}{3in}{\includegraphics{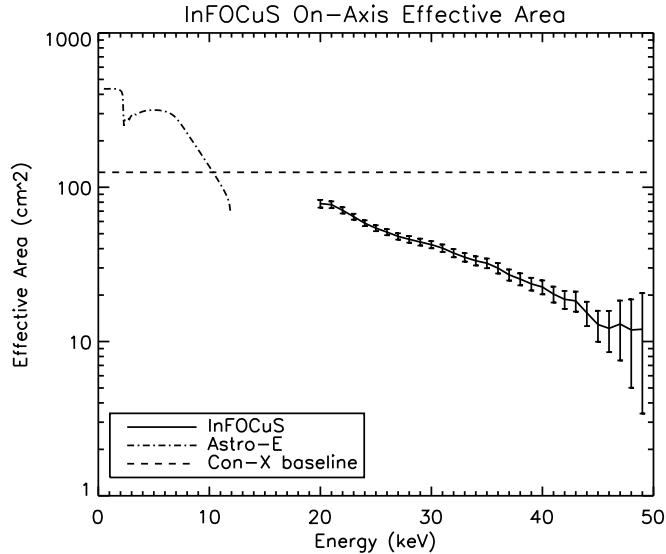}}
\end{center}
\caption{Effective area of the first \infocus\ multilayer mirror.
These data come from a raster scan of a collimated X-ray tube onto the
mirror and focal plane that was performed in the lab before
flight\cite{owens}.}
\label{effarea}
\end{figure}

The detectors on \infocus\ have flown on two balloon flights: a June
2000 flight of only the focal plane, and the first science flight of
the entire telescope in July 2001.  The first science flight in July
2001 included only one mirror and detector in the low energy band;
future flights will add more mirrors and detectors as they are
completed.  This paper concentrates on the detector performance;
mirror results can be found in\cite{b,og,ok,owens}.  For reference,
the effective area of the first \infocus\ 20--40\,keV multilayer
mirror is given in Figure~\ref{effarea}.


\section{FOCAL PLANE}

\begin{figure}
\begin{center}
\rotatebox{270}{\resizebox{!}{4.5in}{\includegraphics{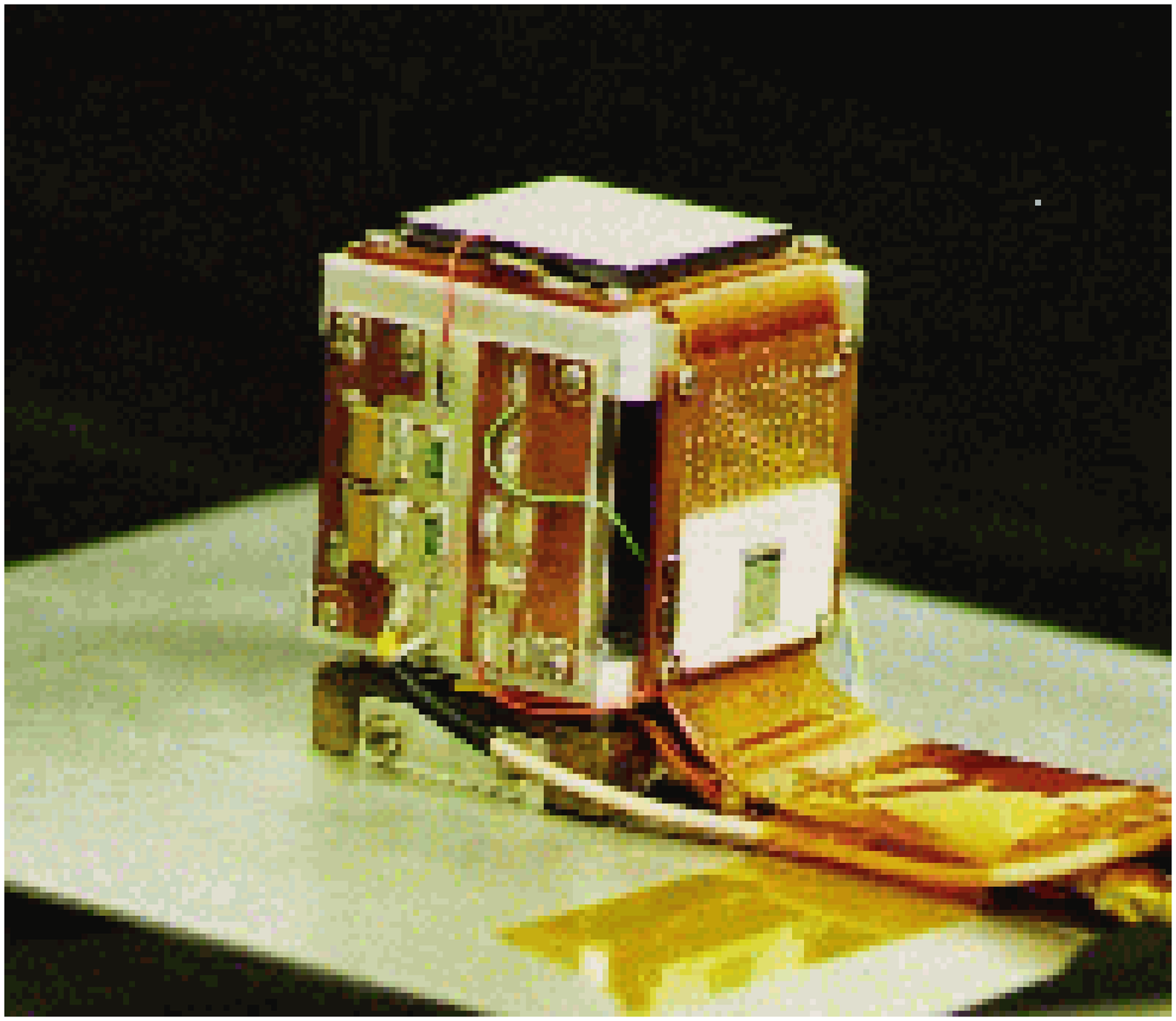}}}
\end{center}
\caption{The \infocus\ detector cube assembly.  The 2.7\,cm square CZT
detector is the thin slab on top of the cube.  The bias HV supply is
filtered by the components on the nearest side panel and connected to
the detector by the small wire leading to the top of the cube.  The
XA-1~ASIC is visible mounted on a white ceramic board on the right
panel of the cube just above the ribbon cable leading to the readout
electronics.  The long trace length between the CZT and the ASIC adds
to the input capacitance and is the leading contributor to the
electronics noise that limits the energy resolution.}
\label{cubepic}
\end{figure}

The active components of the \infocus\ focal plane consist of a
2.7\,cm $\times$ 2.7\,cm $\times$ 0.2\,cm piece of CZT, and a 3\,cm
thick CsI anticoincidence shield.  The CZT detector is configured with
a planar Pt contact on one of the large faces, and a 12 $\times$ 12
segmented array of contacts on the opposite face delineating the
detector pixels.  The pixels are 2.0\,mm square, and are placed on
2.1\,mm centers.  The detector is mounted on the top of a 4\,cm
cube of aluminum that serves as a base for the detector and its
associated electronics, and as a coldfinger to connect the detector to
a thermoelectric cooler and liquid heat exchanger.  A photograph of
the detector cube assembly is shown in Figure~\ref{cubepic}.

The detector is connected with conductive epoxy dots under each pixel
to traces on a PC board that route the detector signals through a
semi-rigid flex cable and onto a separate circuit board on the side of
the cube.  This circuit board contains the decoupling capacitors and
bleed off resistors for each pixel, a ceramic fan-in assembly, and an
XA-1 ASIC.  The filter circuitry for the detector bias voltage is
located on a separate side panel of the cube.  The detector bias
voltage is applied so that the top planar contact acts as the cathode
and the pixels as anodes.  The operating voltage is typically $-200$
volts applied to the cathode, leaving the pixels and the ASIC inputs
at ground potential.

The XA-1 ASIC we use is the same one used in the \textsl{Swift} BAT
detector, and accepts 128 input channels.  (Four pixels in each corner
of the 12 $\times$ 12 detector pixel array are left unconnected in
order to arrive at 128 input signals for the XA-1.) The ASIC contains
a charge sensitive preamplifier and shaping circuit for each pixel,
and allows the setting of a threshold for event detection.  The ASIC
also contains a pulser circuit that can be routed to the inputs in
order to test the response of the system. 

\begin{figure}
\begin{center}
\rotatebox{270}{\resizebox{!}{4.5in}{\includegraphics{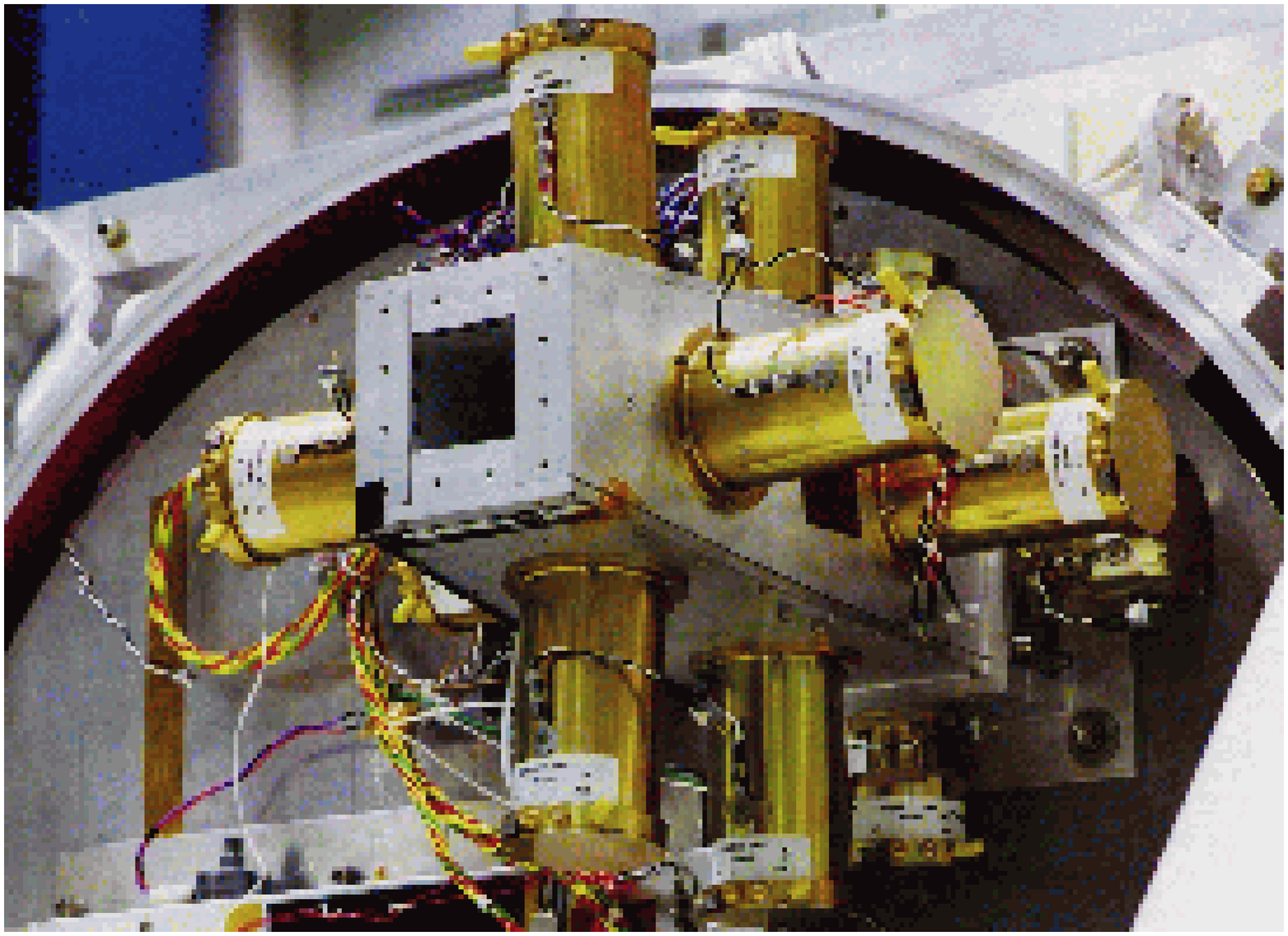}}}
\end{center}
\caption{The \infocus\ active shield.  The detector cube sits in the
bottom of the square well formed by the 3~cm thick CsI scintillator.  The
cylinders projecting from the shield contain the photomultiplier tubes
used to form the veto signal.}
\label{shield}
\end{figure}

We place the detector cube assembly in the bottom of a well of 3.0\,cm
thick CsI that acts as an active shield to reduce the background from
particles and photons not incident along the mirror focal direction.
We are able to set an operating threshold of 15\,keV for the shield,
and operate it as an anti-coincidence veto for signals recorded in the
detector. The detector sits 32\,cm behind the opening on the front
surface of the shield, and sees a shield opening angle of
8.1$^{\circ}$. The 15\,keV threshold is sufficient to reduce
contamination from high energy background components.
Figure~\ref{shield} shows a photograph of the \infocus\ shield
assembly.

\section{DETECTOR TEST RESULTS}

The CZT detector program at Goddard has tested several designs for a
CZT-ASIC combination.  All of them are designed around the same size
piece of CZT, 2.7\,cm$\times$2.7\,cm$\times$0.2\,cm.  We were able to
achieve these large dimensions for our detector with the aid of IR
scanning and X-ray screening techniques that allowed us to examine CZT
wafers in order to find large homogeneous pieces without
defects\cite{bp99,bp01}.

\begin{figure}
\begin{center}
\resizebox{.48\textwidth}{!}{\rotatebox{0}{\includegraphics{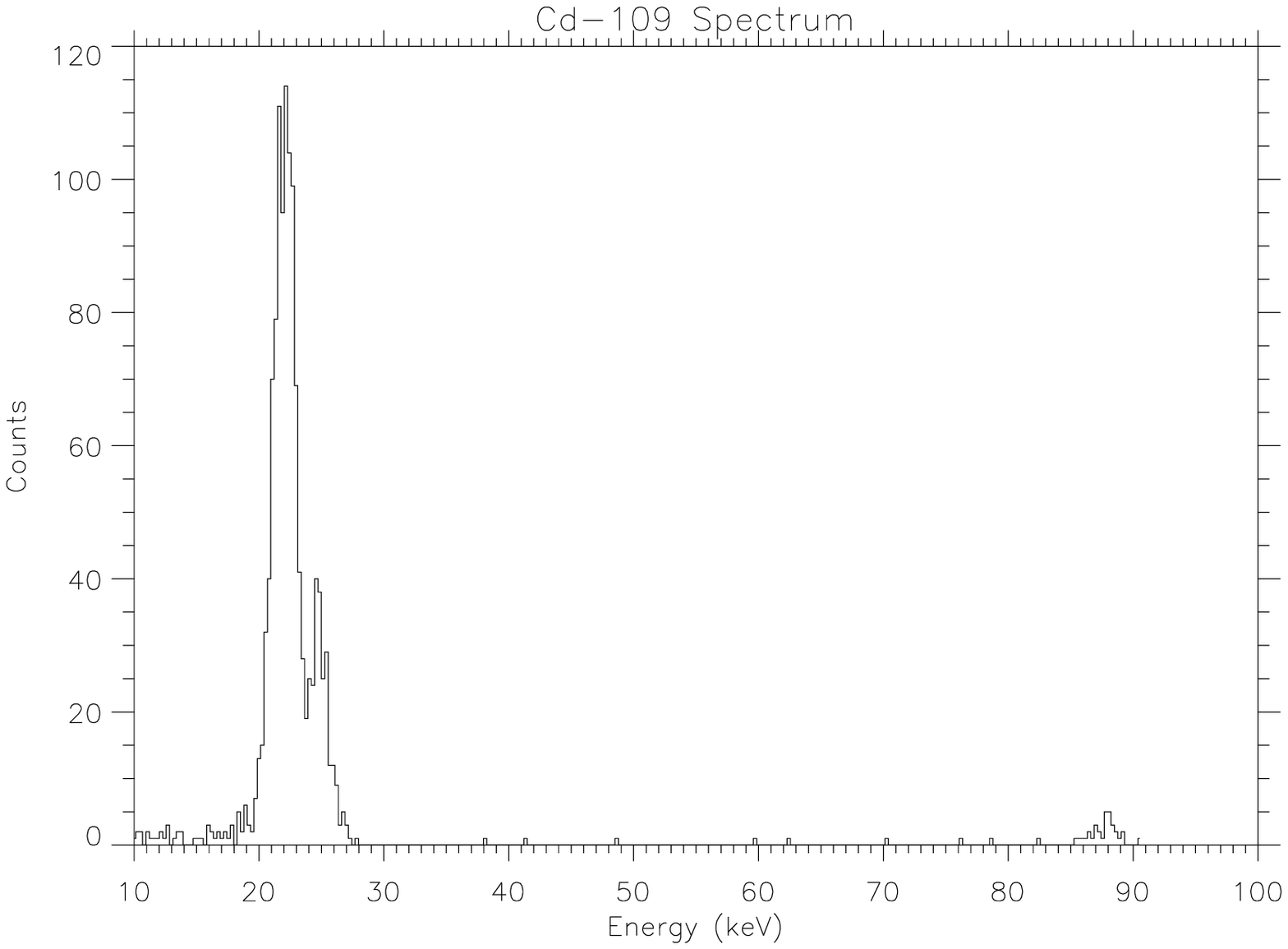}}}
\resizebox{.48\textwidth}{!}{\rotatebox{90}{\includegraphics{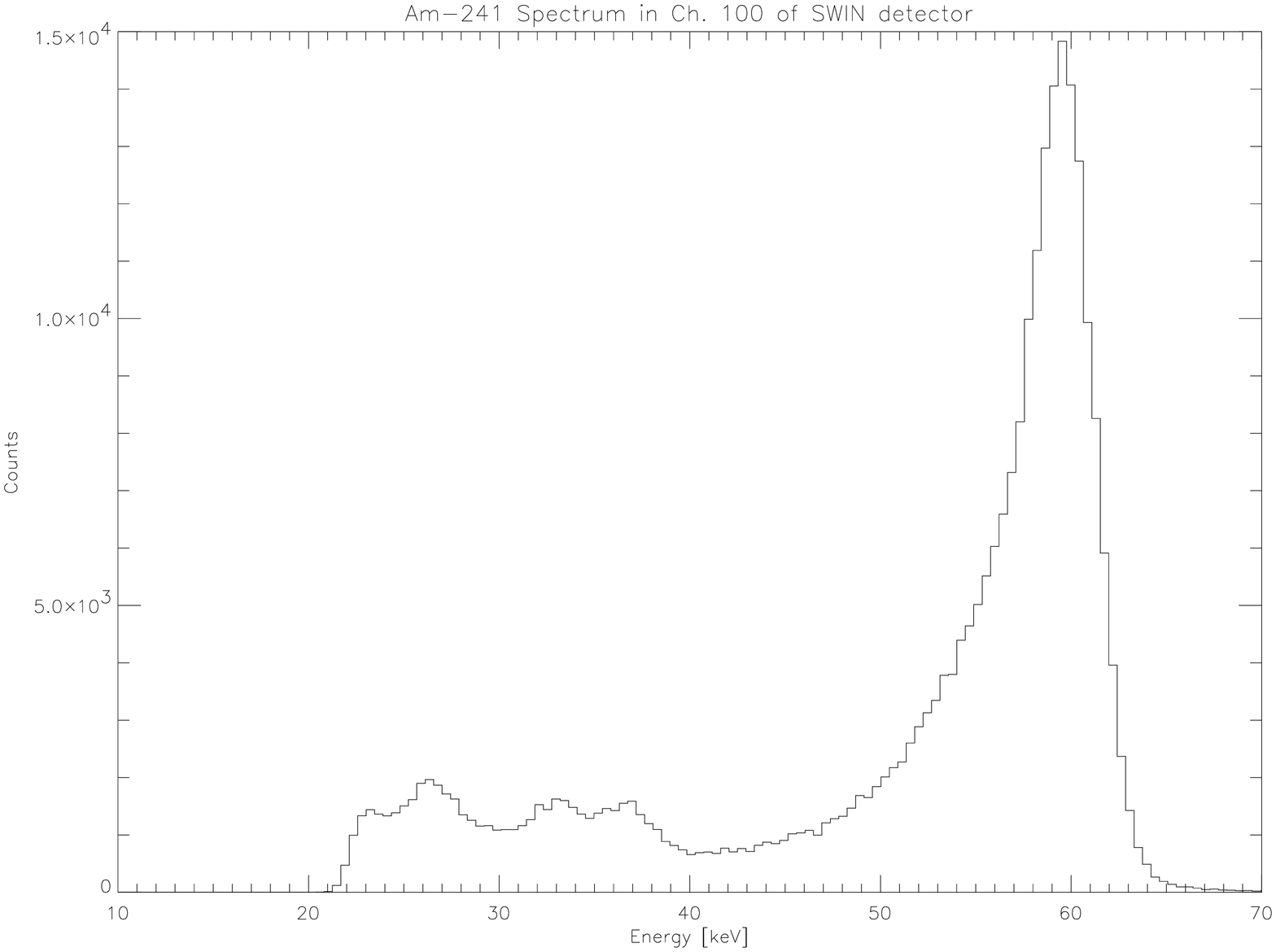}}}
\end{center}
\caption{Single pixel CZT spectra.  The left panel is a $^{109}$Cd
spectrum from the University of Arizona detector.  The source was
collimated to shine onto the center of a 380\,$\mu$m pixel.  The
detector was 2\,mm thick, and operated at 0$^\circ$\,C with a bias of
$-200$ volts.  The resolution at the 22\,keV line is 2.3\,keV FWHM.
The right panel is an $^{241}$Am spectrum from the flight SWIN
detector.  The detector was flood illuminated, and a representative
pixel chosen.  The resolution at the 60\,keV line is 4.8\,kev FWHM.}
\label{cd109}
\end{figure}

The first generation detector system tested at Goddard was based on a
University of Arizona ASIC designed for medical imaging.  This ASIC
utilized an integrating amplifier mechanism which read out every pixel
each readout cycle, had a 64$\times$64 array of 380\,$\mu$m pixels,
and was constructed so that the CZT detector could be directly
connected to the ASIC using indium bump bonding.  With this detector
we achieved an energy resolution of 2.3\,keV FWHM for the 22\,keV
$^{109}$Cd line (Figure~\ref{cd109}),
but the ASIC readout mechanism had an unacceptable deadtime for
astronomical applications.  The main advantage of the UA detector was
the excellent energy resolution that came from the small pixel
effect\footnote{The small pixel effect is a geometry dependent effect
present in pixellated planar detectors that have pixel sizes small
compared to the detector thickness.  The effect allows for increased
spectral resolution because most of the signal is generated by high
mobility electrons instead of low mobility holes.} and the low input
capacitance inherent to the bump bonding technique.

The second generation detector developed at Goddard was based on an
ASIC that used a sparse readout mechanism.  Like the UA detector, it
also was an integrating ASIC meant to be directly bump bonded to a CZT
detector with small pixels.  The sparse readout mechanism was designed
to only read out those pixels with events by comparing the pixel
signals to a set of hold capacitors storing a bias frame.
Unfortunately, the Mitel foundry 3\,$\mu$m process used in the ASIC
construction was unable to produce a matched set of hold capacitors
for each pixel that allowed a suitable triggering mechanism to be
implemented. More troubling, the digital switching signals in the ASIC
easily coupled through the CZT into the inputs, severly increasing
noise and reducing the energy resolution achievable with this design.
The switching signals also coupled onto the top side cathode signal,
degrading it and preventing us from using it as an event trigger.

The third detector design tested at Goddard was a fallback plan based
on the design for the \textsl{Swift} BAT CZT detector, and was called
the SWIN detector (for \textsl{Swift}-\infocus).  This is the detector
design that was flown on the balloon flights.  The XA-1 is not an
integrating ASIC, and uses the more conventional design of a charge
sensitive preamplifier and shaping stages before a discriminator to
detect events above a preset threshold.  The SWIN detector also
differs from the previous detectors in that the CZT is not directly
bump bonded to the ASIC; in SWIN, the signal is routed from the CZT to
the ASIC via traces on a circuit board and a fan-in assembly.  Also,
the XA-1 ASIC has only 128 channels, so the pixel size was increased
from 380\,$\mu$m to 2.0\,mm in order to match the number of pixels to
the number of ASIC channels.  A 2.0\,mm pixel still oversamples the
4.0\,mm PSF of the mirror, but no longer benefits from the small pixel
effect.

\begin{figure}
\begin{center}
\resizebox{!}{7in}{\rotatebox{0}{\includegraphics{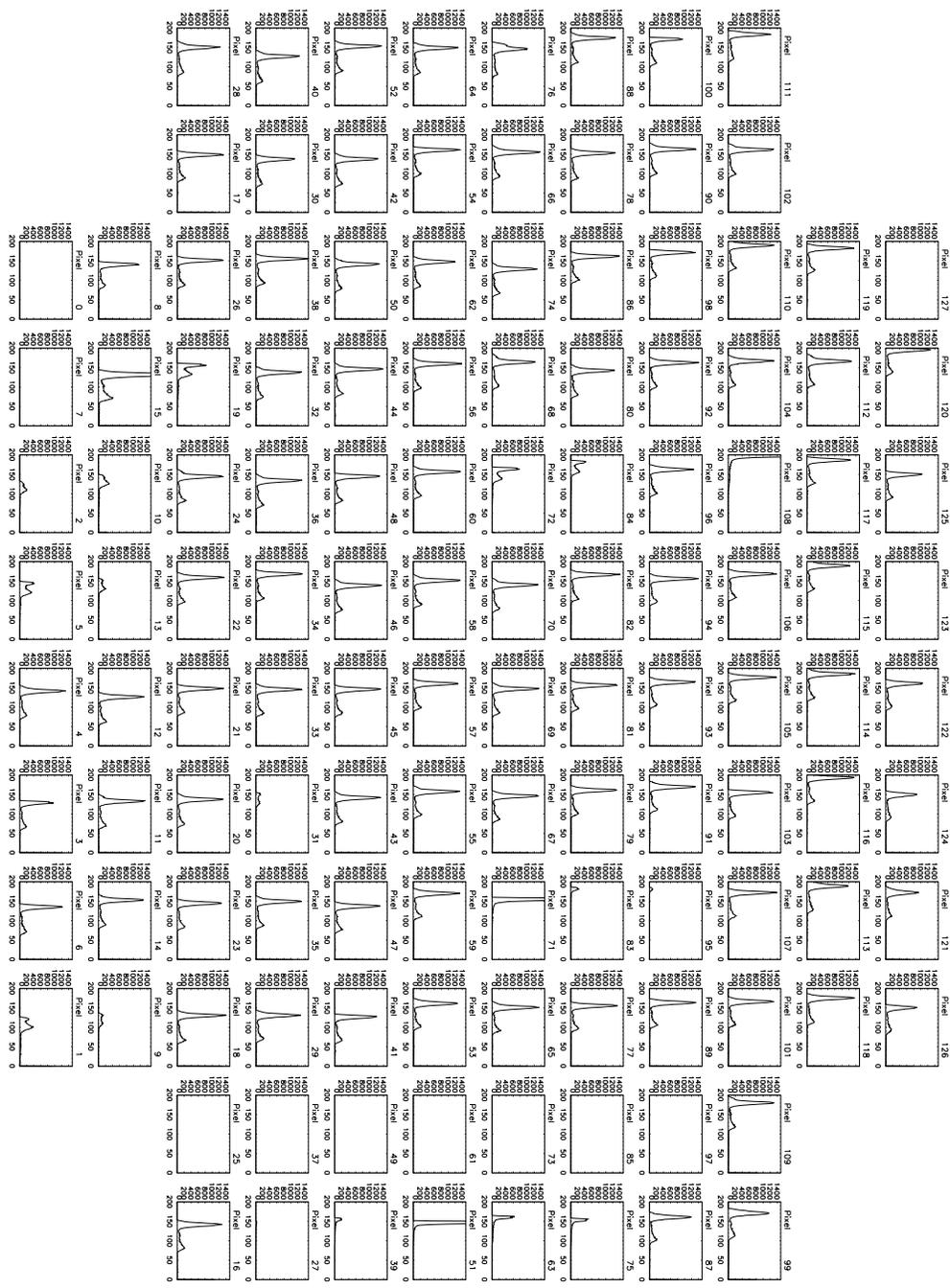}}}
\end{center}
\caption{$^{133}$Ba spectra in the SWIN detector.  There is one graph
for each pixel of the detector.  The vertical axes are counts, and the
horizontal axes are detector channel number and are related to the
event energy.  This data was produced by flood illuminating the
detector with the $^{133}$Ba source, and are similar to that used to
calibrate the gain and offset of the detector.  The data organized in
this form are an end to end demonstration of the operation of the
entire focal plane and make it easy to locate bad pixels to be
electronically masked off in flight.  (The pixels without spectra in
this figure are pixels that have already been masked). Bad pixels were
usually the result of shorting caused by a spreading of the conductive
epoxy used to connect the detector to the carrier board.}
\label{spectra}
\end{figure}

We measured the SWIN detector for uniformity of response among the
pixels and for gain and offset determination by flood illuminating the
detector with a $^{133}$Ba source.  Test data for the SWIN detector
taken under flight-like conditions ($T$ = 10$^{\circ}$\,C and bias
voltage = $-200$ volts) are given in Figure~\ref{spectra}.
Flood illumination data show that the SWIN detector response is rather
uniform across the face of the CZT.  About 10\% of the pixels suffer
from low gain or very high noise.  The position of these pixels is not
coincident with any known defects in the crystal found by imaging with
an IR scan.  Since most of these pixels are adjacent to each other on
the detector, we conclude that they are shorted pixels caused by
spreading of the conductive epoxy dots joining the CZT to the carrier
board.  Tests with a collimated source confirm this conclusion.  This
problem does not significantly affect the telescope performance
because the central part of the pixel array at the focus of the mirror
is relatively free of shorted pixels.

\begin{figure}
\begin{center}
\resizebox{!}{3in}{\rotatebox{90}{\includegraphics{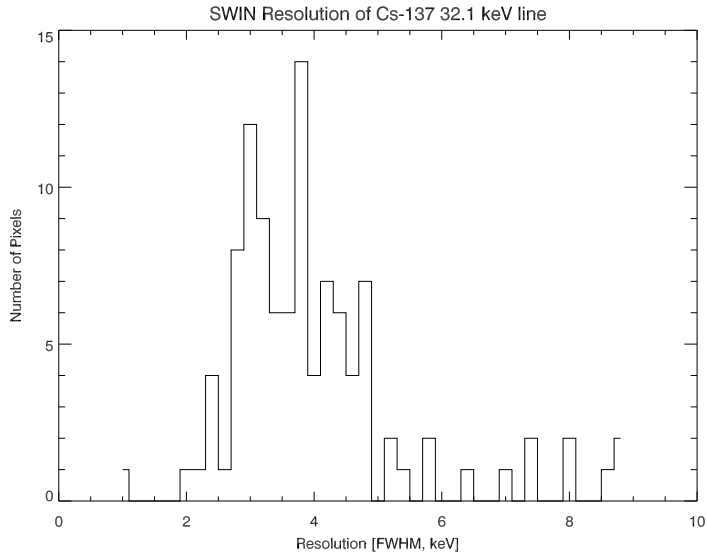}}}
\end{center}
\caption{A histogram of the FWHM energy resolution of the SWIN detector
pixels at the $^{137}$Cs 32.1\,keV line.}
\label{reshists}
\end{figure}

The spectral resolution of an average pixel in the SWIN detector under
flood illumination at the 60\,keV $^{241}$Am line is 4.8\,keV
(Figure~\ref{cd109}). The overall leakage curent of the entire
detector under flight conditions is about 20\,nA.  The measured
resolution of the SWIN detector at the $^{137}$Cs 32.1\,keV line is
4.0\,keV FWHM (see Figure~\ref{reshists}).

With these spectral resolution measurements at two different energies,
we can begin to decompose the line spread function into its different
components.  To first order, the line spread function is composed of
an energy independent part and a component that is linearly dependent
on energy:
\[\sigma^2 = \sigma_n^2 + \sigma_{cc}^2,\]
where $\sigma$ is the measured energy resolution, $\sigma_n$ is the
constant term, and $\sigma_{cc}$ is the energy dependent term.  The
constant term, $\sigma_n$, is primarily due to noise (such as leakage
current), and the energy dependent term, $\sigma_{cc}$, depends
heavily on the charge collection properties of the CZT material.  If
we take $\sigma_{cc} = \alpha E$, then we have
\[\sigma^2 = \sigma_n^2 + \alpha^2E^2,\]
an equation in the two unknowns $\sigma_n$ and $\alpha$.  Using our
two measured spectral resolutions at 32.1\,keV and 59.5\,keV and
eliminating $\alpha$, we obtain $\sigma_n = 3.6$\,keV.  Solving for
$\sigma_{cc}$, we get $\sigma_{cc30} = 1.7$\,keV at 32.1\,keV and
$\sigma_{cc60} = 3.2$\,keV at 59.5\,keV.  These numbers indicate that
flat spectrum noise is the dominant contribution to the line spread
function in our energy range of 20--40\,keV.  Incomplete charge
collection in the CZT detector becomes important at higher energies,
but is only a secondary factor at 30\,keV.


In order to investigate the noise contribution to the line spread
function, we used the internal XA-1 pulser circuit connected to the
inputs of the ASIC.  Pulser measurements of the ASIC alone show a line
spread function of less than 1\,keV, indicating that noise generated
within the ASIC electronics is not the dominant component of the noise
contribution.  Pulser resolution measurements with the CZT detector
connected to the ASIC inputs are consistent with measurements taken
with a radioactive source, indicating that the dominant source of
noise contributing to the energy independent noise term is external to
the ASIC and comes from the detector and its circuitry.  We varied the
temperature of the detector/ASIC and the bias voltage to investigate
the effects of leakage current on the line spread function and find
that the leakage current component of the noise is not a dominant
contribution.  The most likely cause of the high noise term is the
high input capacitance caused by the long lead lengths connecting the
pixels to the ASIC.

\section{FLIGHT RESULTS}
\label{flight_results}

\subsection{The August 2000 Focal Plane Test Flight}

In August 2000, we took \infocus\ to Palestine, Texas to fly the
detector at the National Scientific Balloon Facility.  The main
purpose of the flight was to test our detector at float altitudes to
determine the count rate from the particle background, and the
effectiveness of our anti-coincidence shield system.  This measurement
was a key parameter in the determination of the sensitivity of the
instrument.  No mirror was present, and no astronomical objects were
observed. Also flying on this test flight were two BAT detectors from
the \textsl{Swift} mission that were being checked for similar reasons.
Another main goal of the flight was to check other flight systems in a
realistic environment.

The balloon was launched on August 29, flew for seven hours at
118,000\,ft, and was successful in measuring a relatively flat CZT
background in flight of $(7.1\pm 3.5) \times 10^{-4}\ \mathrm{cts}\
\mathrm{sec}^{-1}\ \mathrm{cm}^{-2}\ \mathrm{keV}^{-1}$.

\subsection{July 2001 Science Flight}

In July, 2001 \infocus\ flew its first science flight from with a
complete mirror and detector system from Palestine, Texas.  This
flight achieved three hours at float altitude and allowed \infocus\ to
become the first telescope to utilize a multilayer mirror and detect
an astronomical source using CZT detectors.

\subsubsection{Cygnus X-1 results}
\label{sec:cygflux}

\begin{figure}
S\begin{center}
\resizebox{!}{3in}{\rotatebox{0}{\includegraphics{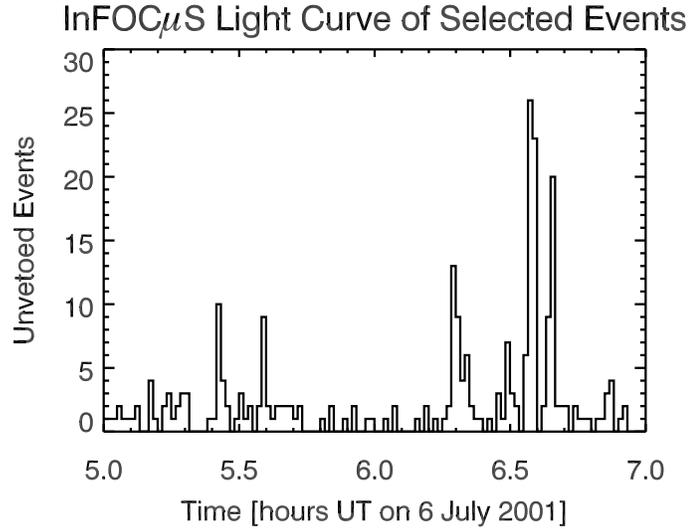}}}
\end{center}
\caption{the \infocus\ Cyg X-1 light curve.  The data is binned in one
minute intervals.  The low level of background is evident, as well as
peaks when the pointing coincided with the position of the source.  If
the pointing had been on target at all times, we would have received
about 80 counts per minute from Cyg~X-1.}
\label{cyg_lc}
\end{figure}

During the July 2001 science flight, the pointing system failed to
achieve the necessary one arcminute pointing to track a source.
\infocus\ concentrated on obtaining as many photons as possible from
the bright source Cyg~X-1.  Analysis of the attitude data after the
flight showed stochastic long time scale 30--45~arcminute pointing
errors probably caused by turbulence and wind shifts in the upper
atmosphere.  Figure~\ref{cyg_lc}
shows the lightcurve for a two hour observation of
Cyg~X-1, and indicates several high count rate periods when we
detected Cyg~X-1 when the pointing was momentarily good.

\begin{figure}
\begin{center}
\resizebox{.48\textwidth}{!}{\rotatebox{0}{\includegraphics{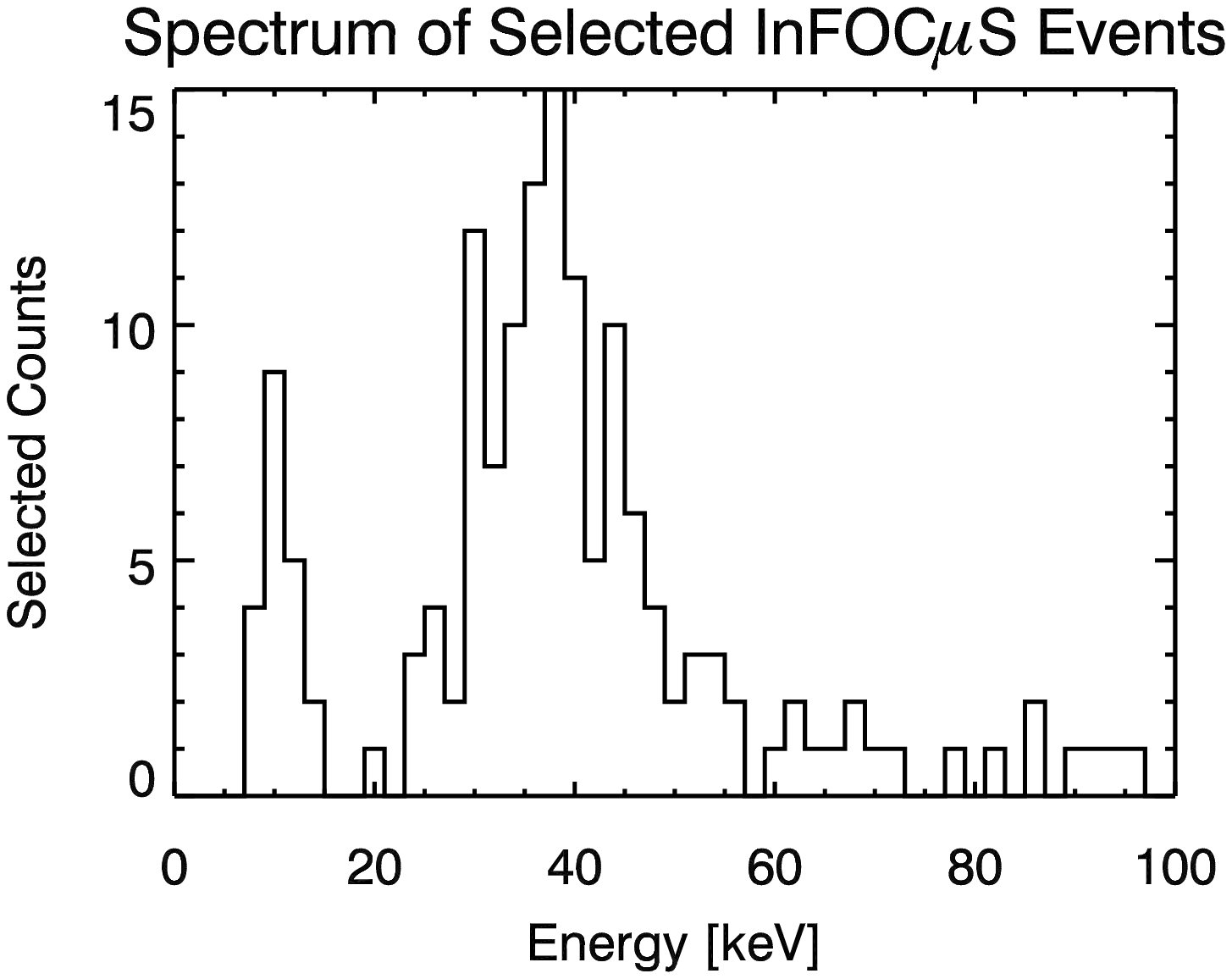}}}
\resizebox{.48\textwidth}{!}{\rotatebox{0}{\includegraphics{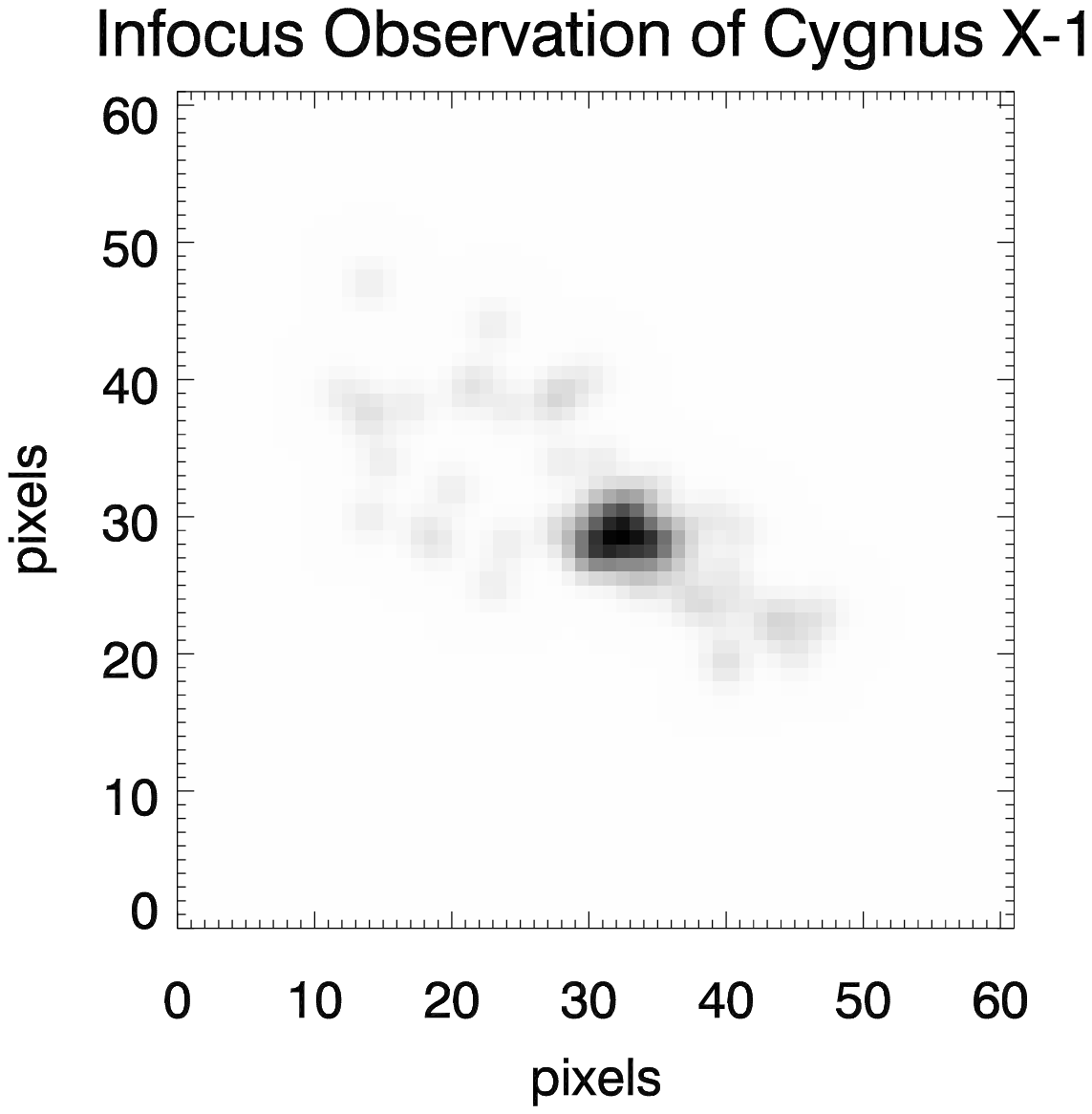}}}
\end{center}
\caption{The \infocus\ astronomical results.  The left panel shows the
Cyg X-1 spectrum.  The peak centered at about 30\,keV matches very
well to our pre-flight predictions of the Cyg~X-1 spectrum.  The small
peak below 20\,keV results from noise below the detector
threshold. The energy scale could only be roughly calibrated to
laboratory data because of the failure of the in-flight pulser
calibration mechanism.  The right panel shows the \infocus\ Cyg X-1
image.  Each detector pixel is 54~arcseconds.  The image was made by
using gyro and star camera data to determine the aspect of the
telescope for each photon detected within the bandpass of the mirror
during periods when the Cyg~X-1 count rate was high.  The photon map
was divided by the exposure map and then convolved with the
2~arcminute HPD of the telescope PSF.  The peak of the exposure map
indicates about 80 seconds on the location of Cygnus~X-1.}
\label{cygnus}
\end{figure}

Figure~\ref{cygnus}
shows the spectrum extracted from these high count rate times and
confirms the detection of Cyg~X-1.  The peak indicated in the spectrum
between 30--50\,keV falls in the approximate location of the mirror
bandpass, trails off at low energy as would a continuum source
attenuated by the atmosphere, and also falls at high energies as the
effective area of the mirror decreases.  The apparent mismatch between
the 20--60\,keV width of the peak and the 20--40\,keV mirror bandpass
is due to ASIC offset drift that could not be calibrated because of a
defective inflight pulser system.  The actual calibration used is
based on the best available ground calibration with radioactive
sources.  Figure~\ref{cygnus} also shows an image of the photons
detected during these high count rate times derived from post-flight
attitude reconstructed data convolved with the telescope PSF.  

We also used the exposure corrected image to determine the flux of
Cyg~X-1. Starting from the exposure corrected image, we subtracted the
background rate to obtain the observed Cyg~X-1 count rate in the
20--40\,keV band.  We then folded the Cyg~X-1 spectrum given by
Doebereiner et al\cite{d} through our model atmosphere
(3.3 g cm$^{-2}$ and an elevation of 75$^{\circ}$) and measured mirror
effective area to obtain the model count rate.  The ratio of the count
rates multiplied by the model flux gives us our observed flux of
$4.9\times10^{-9}\ \mathrm{ergs}\ \mathrm{sec}^{-1}\
\mathrm{cm}^{-2}$.  This value is in good agreement with the
Doebereiner \textsl{HEXE} result of $6.48\times10^{-9}\ \mathrm{ergs}\
\mathrm{sec}^{-1}\ \mathrm{cm}^{-2}$ given the known variability of
Cyg~X-1.

\subsubsection{Inflight background}

\begin{figure}
\begin{center}
\resizebox{!}{3in}{\rotatebox{90}{\includegraphics{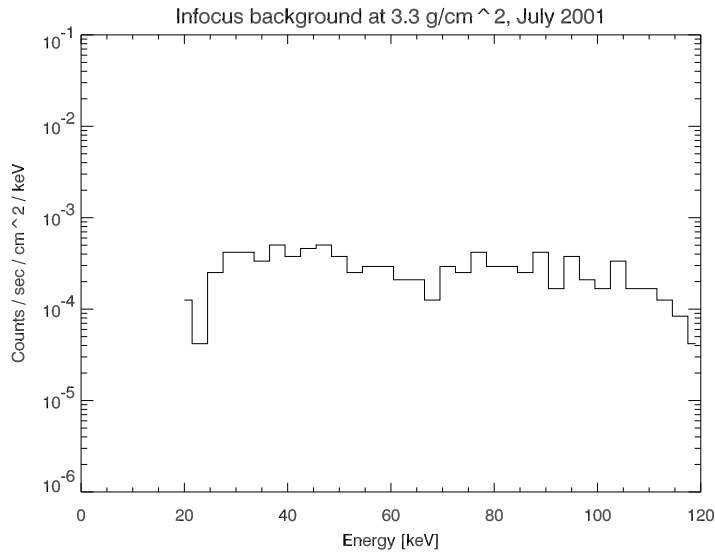}}}
\end{center}
\caption{\infocus\ background spectrum.  This spectrum was measured
using the \infocus\ SWIN detector during the science flight of July
2001 from Palestine, Texas during times when we were not observing an
astrophysical source.  The balloon altitude was 130,000\,ft and the
shield threshold was set at 15\,keV.  The spectrum is well
characterized by the the value $(2.7\pm 1.2)\times 10^{-4}$ cts
sec$^{-1}$ cm$^{-2}$ keV$^{-1}$ in the band from 20--120\,keV.}
\label{bkg_spec}
\end{figure}

The other principal goal of the flight was to measure the background
for the CZT detectors in order to determine the sensitivity of
\infocus\ in future flights, to verify the measurement made in the
test flight of 2000, and to demonstrate the suitability of our active
CsI shield.  Figure~\ref{bkg_spec}
shows the \infocus\ detector background determined in July 2001.  The
measured background from a two hour observation is rather flat from
20--120\,keV, and lies at $(2.7\pm 1.2)\times 10^{-4}$ cts sec$^{-1}$
cm$^{-2}$ keV$^{-1}$.  This number was determined by taking the
spectrum in Figure~\ref{bkg_spec} and computing the average and
variance of the background level in all the 3\,keV wide bins between
20--120\,keV.  The somewhat high uncertainty is mostly statistical and
comes from the fact that even with a two hour observation we have very
few photons ($\sim10$) in each bin.

This measurement was taken at float altitude of 130,000\,ft without
using any depth determination techniques to further reduce the
background.  This measurement gives a lower background than the test
flight measurement, but we are more confident of the lower value
because the test flight background spectrum was slightly contaminated
by interference noise in some of the pixels.  We have carefully
measured the deadtime in the detector, and find that the largest
contribution (from the shield veto) gives less than 1.5\% deadtime.
The correct measurement of the Cygnus~X-1 flux
(Section~\ref{sec:cygflux}) verifies that the signal chain is
operating correctly and does not suffer from any deadtime effects that
could affect the background level.

Rothschild\cite{roth} and Slavis\cite{slavis} show that the dominant
CZT background components are uniformly distributed in the volume of
the CZT detector and that depth determination techniques can reduce
the background level by a factor of seven.  If we apply their factor
of seven to our uncorrected measured results, the \infocus\ background
is the lowest CZT background yet measured.  These results indicate
that a heavy active shield operated with a low threshold in
conjunction with a depth determination algorithm are both necessary to
achieve the lowest possible CZT background levels and the best
instrument sensitivity.

\section{CONCLUSIONS AND FUTURE PLANS}

The \infocus\ telescope has demonstrated the effectiveness of a CZT
detector coupled to a multilayer mirror for hard X-ray astronomy.  The
moderate resolution SWIN detector has become the first CZT detector to
observe astrophysical photons, and our heavy CsI shielding has enabled
us to achieve a very low in flight background.  The \infocus\
implementation needs improvements in the pointing system before it can
do high precision astronomy, but the new technology behind the mirror
and detector have proven sound.  Even with an on target observation
time of only about a minute, we have detected the astrophysical source
Cyg~X-1 with very high significance.

\subsection{Detector Improvements}

Our laboratory and flight experience with the SWIN CZT detector and
the \infocus\ telescope have given us valuable experience to direct
future improvements.  The most important goal is to improve the
spectral resolution of the detector.  Our plans for the next
generation \infocus\ detector will have an improved layout that will
reduce input capacitance by shortening the lead length between the CZT and
ASIC.  Reducing the pixel size and adding steering electrodes will
improve charge collection, also improving the energy resolution.
Another possibility for reducing the input capacitance is to remove
the decoupling capacitors between the CZT and ASIC.  This requires a
very low leakage current, but this can be achieved by running the
detector much colder.  Takahashi et al.\cite{tak} have shown improved
energy resolution in CZT detectors by operating at a very high voltage
in order to improve charge collection.  This also requires very low
temperatures in order to minimize the leakage current, but our
detector temperature regulation system is able to go down to
$-25^{\circ}$ without modification, and colder with some minor
hardware changes. 

New ASICs are another important step towards improving energy
resolution.  We are currently designing a custom ASIC specifically for
our detector configuration that should not be as sensitive to input
capacitance effects.  Also, a newer generation of XA-1 ASIC has been
developed with much improved temperature stability that will
significantly reduce our susceptibility to offset drifts.  This ASIC
is being used in the \textsl{Swift} BAT flight detectors, and will be
tested for future \infocus\ detectors.

Other groups have shown the importance of using a depth determination
mechanism to achieve the lowest detector background and best
sensitivity.  We have been experimenting with two different
techniques, one based on utilizing the cathode signal and another
based on using the measured charge spreading in neighboring small
pixels to determine the depth of interaction.  The cathode signal
method is simpler to operate but more difficult to implement since the
cathode signal has an inherently worse resolution because most of its
signal is generated by holes.  A possible solution to this problem is
to segment the cathode electrode into strips, lowering the capacitance
and improving the resolution so that the cathode signal can be used.

Dan Marks has shown in his Ph.D. thesis\cite{dm} that determining the
photon depth of interaction by observing the signal in the neighboring
pixels is possible.  However, this is also difficult in practice
because of the computational complexity, and because it requires the
neighboring pixel signals.  More recent versions of the XA-1 ASIC than
the one used in SWIN have this capability, and we will explore this
method for background rejection.

\subsection{Future Flight Plans}

The most important enhancement to the \infocus\ platform will be to
improve the pointing system to allow long-term stable pointing at the
one arcminute level.  Our flight experience has given us valuable data
on high-altitude winds and turbulence that will be incorporated into a
future design utilizing more robust pointing mechanisms with a larger
dynamic range.

Future flights of \infocus\ will also fly more advanced technology as
it becomes available.  The Nagoya group is working on a second
low-energy multilayer mirror that will have improved performance and
will fly on the next flight.  Future flights are planned to include
the high-energy mirrors for $^{44}$Ti studies of supernovae remnants.
Each of these planned future flights will use an improved CZT detector
based on the knowledge we have gained with the SWIN detector and
incorporating the changes mentioned above.

\section*{ACKNOWLEDGMENTS}
We would like to thank the NSBF flight crew for all their help and for
a successful flight.  We would also like to especially thank Chris
Miller, Steve Derdeyn, Kiran Patel, Steve Snodgrass, Holly Hancock and
all the dedicated Goddard engineers and technicians whose hard work
and countless extra hours made \infocus\ possible.


\end{document}